\documentclass[aip,jmp,amsmath,amssymb,reprint]{revtex4-1}

\usepackage{graphicx}% Include figure files
\usepackage{dcolumn}% Align table columns on decimal point
\usepackage{bm}% bold math
\usepackage[usenames]{color}
\definecolor{cRef1}{rgb}{1.0,0.0,0.0}
\usepackage{booktabs}
\usepackage{ctable}
\usepackage{multirow}

\begin{document}
 
\title[Ion-acoustic shocks with reflected ions]{Ion-acoustic shocks with reflected ions: modeling and PIC simulations}

\author{T. V. Liseykina}
\email{tatyana.tiseykina@uni-rostock.de}
\affiliation{Institut f\"{u}r Physik, Universit\"{a}t Rostock, Rostock, Germany}
\affiliation{Institute of Computational Technologies SD RAS, Novosibirsk, Russia}

\author{G. I. Dudnikova}
\affiliation{Institute of Computational Technologies SD RAS, Novosibirsk, Russia}

\author{V. A. Vshivkov}
\affiliation{Institute of Computational Mathematics and Mathematical Geophysics SD RAS, Novosibirsk, Russia}

\author{M. A. Malkov}
\affiliation{CASS, University of California at San Diego, USA}
\begin{abstract}
Non-relativistic collisionless shock waves are widespread in space and astrophysical plasmas and are known as efficient particle 
accelerators. However, our understanding of collisionless shocks, including their structure and the mechanisms whereby they 
accelerate particles remains incomplete. 
%Microscopically, these shocks are believed to be supported by suitable plasma waves 
%that randomize particle trajectories in lieu of binary collisions. These waves are driven by non-equilibrium components of 
%the plasma such as shock reflected or shock accelerated particles. 
We present here the results of numerical modeling of an 
ion-acoustic collisionless shock based on one-dimensional (1D) kinetic approximation both for electrons and ions with a \textit{real 
mass ratio}. Special emphasis is made on the shock-reflected ions as the main driver of shock dissipation. 
The reflection efficiency, velocity distribution of reflected particles and the shock electrostatic structure are studied 
in terms of the shock parameters. 
%Moreover, we present an analytic solution for ion-acoustic collisionless shock with reflected ions. 
%It extends a classic soliton propagating at the Mach numbers M < M^∗ \sim 1.6 beyond this value at which the soliton reflects the 
%upstream ions. The soliton turns into a shock whose parameters, such as the maximum and the minimum of the potential in its 
%trailing nonlinear wave, are obtained in terms of the number of reflected ions. 
Applications to particle acceleration in geophysical and astrophysical shocks are discussed. 
\end{abstract}
\maketitle

\section{Introduction}
A fundamental problem of dissipation of flows in collisionless plasmas moving faster than all plasma modes available 
has numerous laboratory, astrophysical and geophysical applications. It was realized more than half a century ago 
(see e.g. \cite{Sagdeev,Kennel,Papadopoulos} for reviews) that despite lacking collisions, a shock wave must form in such flows. 
The new applications pose now new challenges to our understanding of collisionless shocks 
structure and their potential to accelerate particles \cite{Blandford_Eicher,Malkov_Drury,Blandford_etc,Bulanov,Haberberger,Fuiza,Adriani}. 
Microscopically, strong shocks can be supported by suitable plasma waves that randomize particle trajectories in lieu of binary collisions. Although these waves are not necessarily 
%may or may not be 
energetically important,
%(i.e. contribute to the Rankine-Hugoniot relation, radiate part of the shock energy etc.), 
in many cases they are driven by shock reflected particles. In this paper we consider relatively weak 
shocks in which the crucial process of ion reflection can be effectively approached from the well understood physics of solitary waves.
%Its relation to classical non-reflecting solitons propagating at Mach numbers strictly limited by $M< M_{\star}=1.6$ 
%(Boltzmann electrons)\cite{Sagdeev} and $M<M_{\star}=3.1$ (trapped electrons)\cite{Gurevich}, is quantified.

The shock at first place results from a nonlinear steepening of a certain mode whose dispersion often limits the steepening. 
For a number of modes \textit{nonlinearity} and \textit{dispersion} can balance each other exactly which results in a \textit{soliton}. 
%Its amplitude obviously controls particle reflection. 
In this paper we consider the most fundamental ion-acoustic solitons with a trailing wave train \cite{Sagdeev,Moiseev}, which are 
the building blocks of collisionless shock waves in nonisothermal plasmas, with the electron temperature much higher than the ion 
temperature,  $T_e \gg T_i.$
Our study is relevant to the electrostatic shock propagation in laser-produced plasmas 
\cite{Fuiza,Mourou,Palmer,MacchiRMP,Romagnani,Sorasio}, especially to the generation of 
monoenergetic ion beams, and to shock--related processes in astrophysical and space plasmas. Among astrophysical 
%and geophysical 
applications of this study we would like to mention the ion injection into the first order Fermi acceleration 
in astrophysical shocks.  The injection process is crucial for
the resolution of such problems as the problem of cosmic ray origin. Overall, the cosmic ray
spectrum is a power-law starting from supra-thermal energies and the injection efficiency controls
the normalization of the spectrum. It ensures a smooth transition between the thermal plasma
and the power-law spectrum at high energies. Although the power-law component is 
%formed by a
%different acceleration mechanism (it is 
due to multiple shock front crossing and not so much due to the reflection from the shock, the particle reflection can also contribute 
to the high energy part of the spectrum, if the shock is oblique. In this case the magnetic field is important. The electrostatic
mechanism is not efficient at high particle energies. However, our purely electrostatic simulations
are relevant to the microphysics of particle reflection off the shock front including the cosmic ray
loaded shocks.

\section{Model and Simulation set-up}
\begin{figure}
\centerline{
\includegraphics[height=0.25\textwidth]{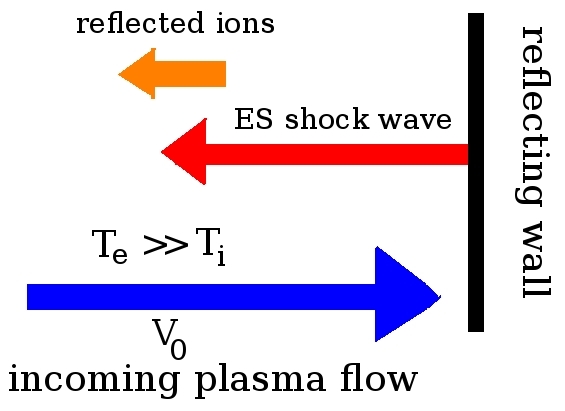}}
\caption{\textit{Scheme of the  set-up for 1D kinetic modeling of electrostatic shock wave}}
\label{fig:set-up}
\end{figure}

In order to address the ions reflected and escaped from the shock, incorporate the ion reflection into the global shock structure and 
investigate its effect on the shock itself we use the kinetic plasma description. The plasma dynamics is governed by 
%the system of equations which consists of 
the Vlasov equations for the distribution functions of the plasma components and Poisson equation for the 
electrostatic field. 
%\[
%\frac{\partial f_{\alpha}}{\partial t}+\vec{v}\frac{\partial
%f_{i,e}} {\partial\vec{r}}+\frac{q_{\alpha}}{m_{\alpha}}\vec{E} \frac{\partial f_{\alpha}}
%{\partial\vec{v}}=0
%\]
%\[
%{\mathrm div}\vec{E}=-\Delta\varphi=4\pi\rho\nonumber
%\]
%\[
%\rho=\sum_{\alpha} q_{\alpha}\int f_{\alpha}d\vec{p}\nonumber
%\]
Numerically, we solve the Vlasov equations using Particle-in-cell (PIC) method \cite{Dawson}, the electrostatic field is obtained by  
direct integration of the charge density, which in turn is calculated from the individual particle positions. For the solution of 
the equations of motion of particles an organic second order in time and space time-reversible numerical algorithm is applied.\\ 
\indent Fig. \ref{fig:set-up} shows the schematic set-up for our 1D simulations. A shock wave is produced by reflection of a high-speed 
electron-ion plasma off a conducting 
wall. The shock forms due to the interaction between the incoming and reflected flows and propagates away from the wall. 
We measure the density of plasma components in units of the unperturbed density $n_0$, coordinates in units of the 
Debye length of the unperturbed plasma flow $\lambda_D=\sqrt{T_e/4\pi e^2 n_0}, $ 
velocity in units of the electron thermal speed $u_0=\sqrt{T_e/m_e},$ and time and electric field in units of 
$1/\omega_{pe}=\sqrt{m_e/4\pi e^2 n_0},$ and $E_0=\sqrt{4\pi n_0 T_e}$ correspondingly.\\
\indent In out PIC simulations, we ensure the high enough resolution to resolve low-density regions, sharp gradients, 
and the dynamics of both electrons and ions by taking at least 100 particles per cell and a spatial resolution 
better than 0.05 $\lambda_D.$ For reference, in all simulations shown the conducting wall is located at $x=0$ and
the plasma flow impinges on it at $t=0.$
Within these settings, the dimensionless input parameter which determines the evolution of the plasma flows is $M_0=V_0/C_s, $ where $V_0$ 
is the speed of unperturbed plasma flow and $C_s=\sqrt{T_e/m_i}$ is the sound speed. 
Moreover, the stability of the soliton and the formation of the shock are proven to be strongly dependent on the velocity distribution of 
ions\cite{MacchiPRE}, thus on the fraction $T_i/T_e$.\\  
% Since the applied treatment include finite 
%ion temperature, the fraction $T_i/T_e$ is a second important input parameter.\\
\indent Note, that this purely electrostatic approach is reasonable for quasi-parallel shocks as it treats scales 
$\lambda_D<l<c/\omega_{pi}\sim r_g^i$ (as $\beta\sim 1$ very often), thus providing an important microphysics ingredient 
for the diffusive shock acceleration as well \cite{DSA,DSA1,DSA2}. The hybrid simulations, broadly used  
in astro-geo-shocks studies, do not resolve these scales since they treat electrons as a fluid.

\section{Simulation results}
\subsection{Cold upstream ions}
\begin{figure*}
\includegraphics[height=0.44\textwidth]{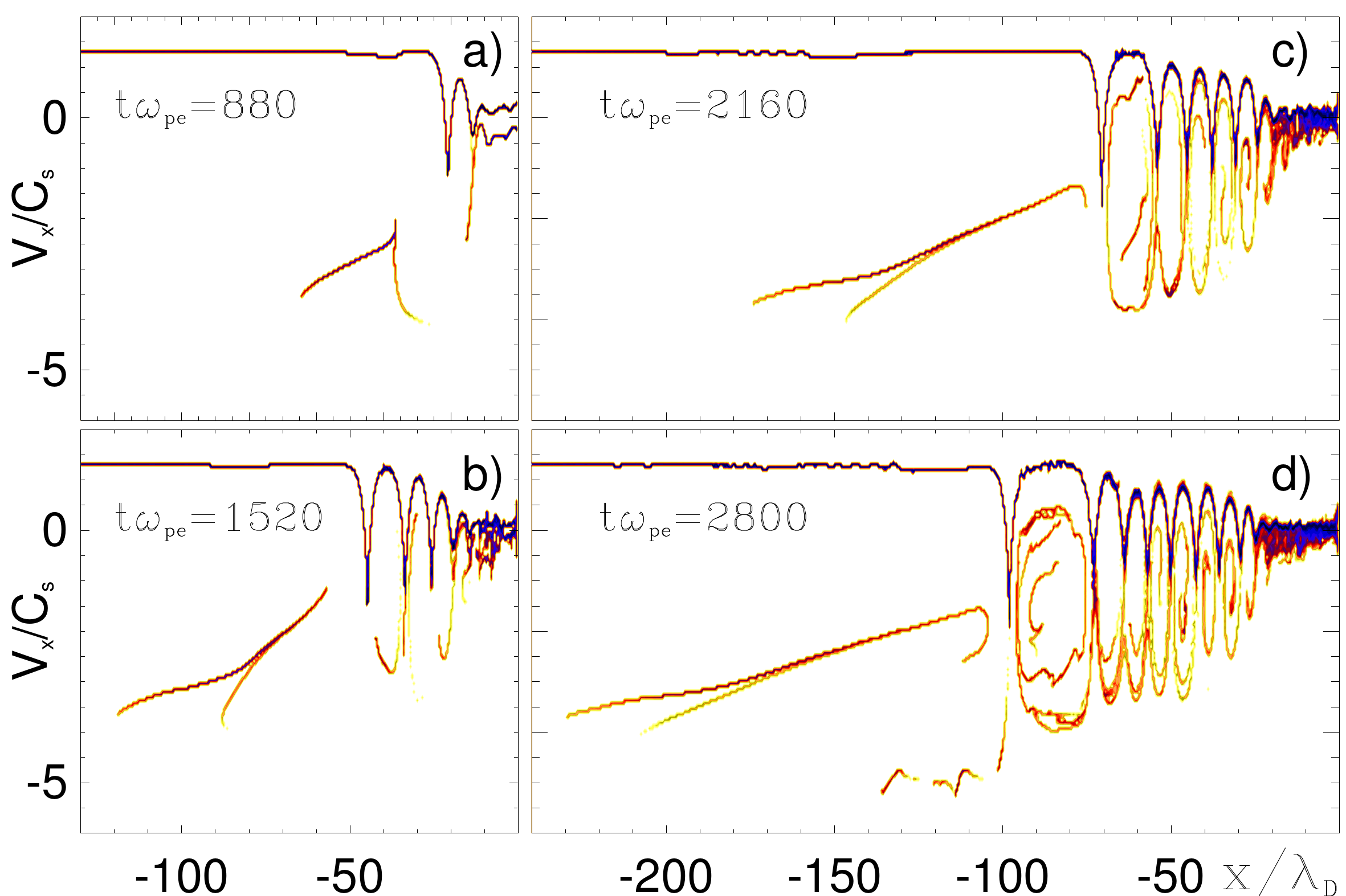}
\includegraphics[width=0.33\textwidth]{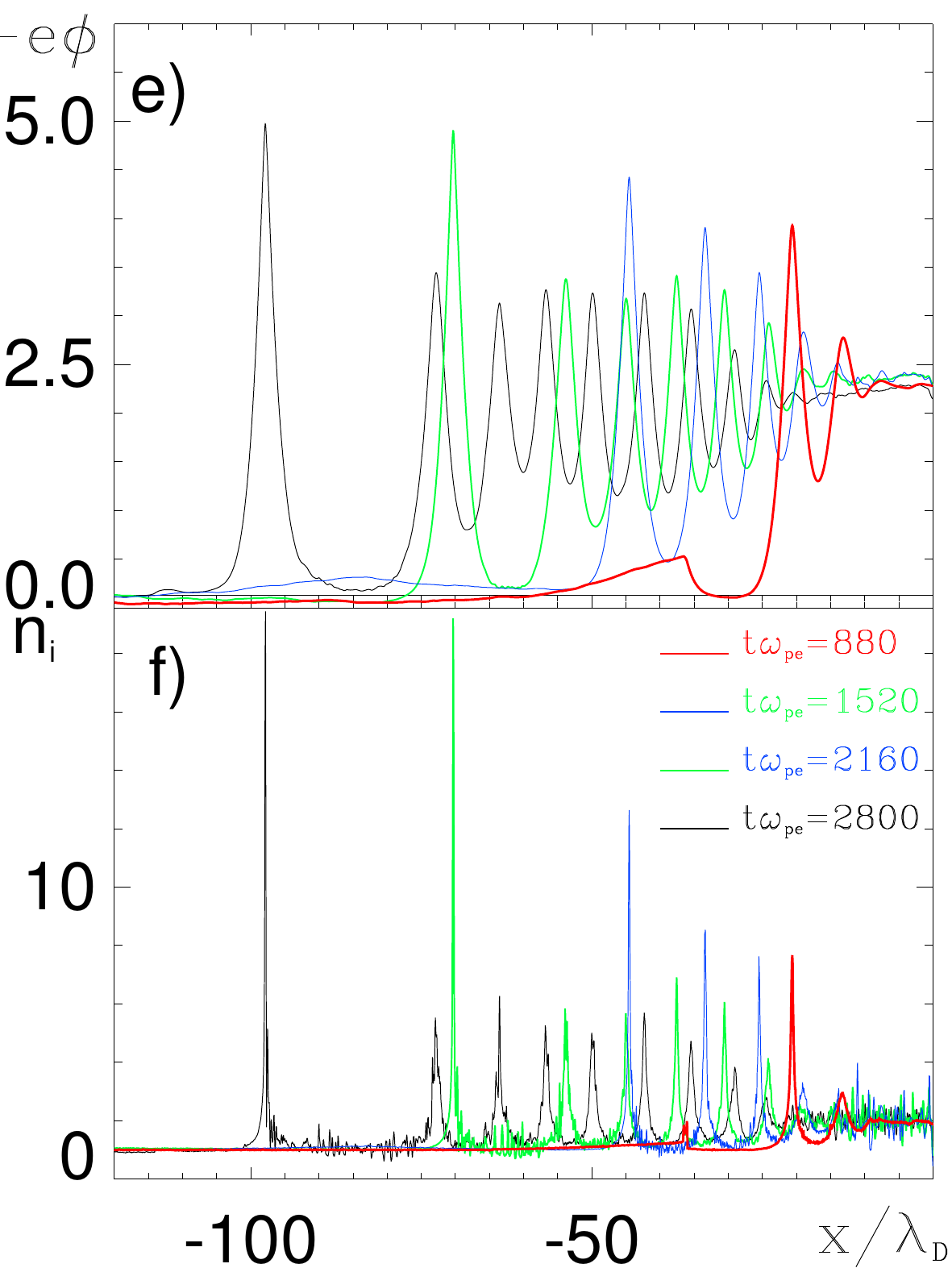}
\caption{\textit{a)--d) -- The distribution of the ions on the $(X, V_x)$ phase plane. e)--f) -- The distributions of the electrostatic potential and 
ion density  at $t=880, 1520, 2160,2800~\omega_{pe}^{-1}$ for $M_0=V_0/C_s=1.3.$ The leading \textit{soliton} 
has a \textit{cusp} structure. It occurs only when the cold ions get reflected.}}
\label{fig:v0=1.3}
\end{figure*}

The evolution of the distribution of ions on phase plane $(x, V_x)$ for $M_0=V_0/C_s=1.3$ is shown in 
Fig. \ref{fig:v0=1.3} a)--d). By $t=200 \omega_{pe}^{-1}$ the flows interacted and the shock is formed. 
The isolated density (and electrostatic potential) structures, Fig. \ref{fig:v0=1.3} e)--f), have the characteristic 
width of several $\lambda_D.$ Ions reflected from the shock are clearly visible on the phase plane. 
%The first outrunning ions are produced by the reflection from the wall. 
In the case of cold upstream ions the ions are reflected at a single point where the potential
reaches its maximum and the electric field has a jump and the potential profile a \textit{cusp}.

\begin{figure*}
\includegraphics[width=0.55\textwidth]{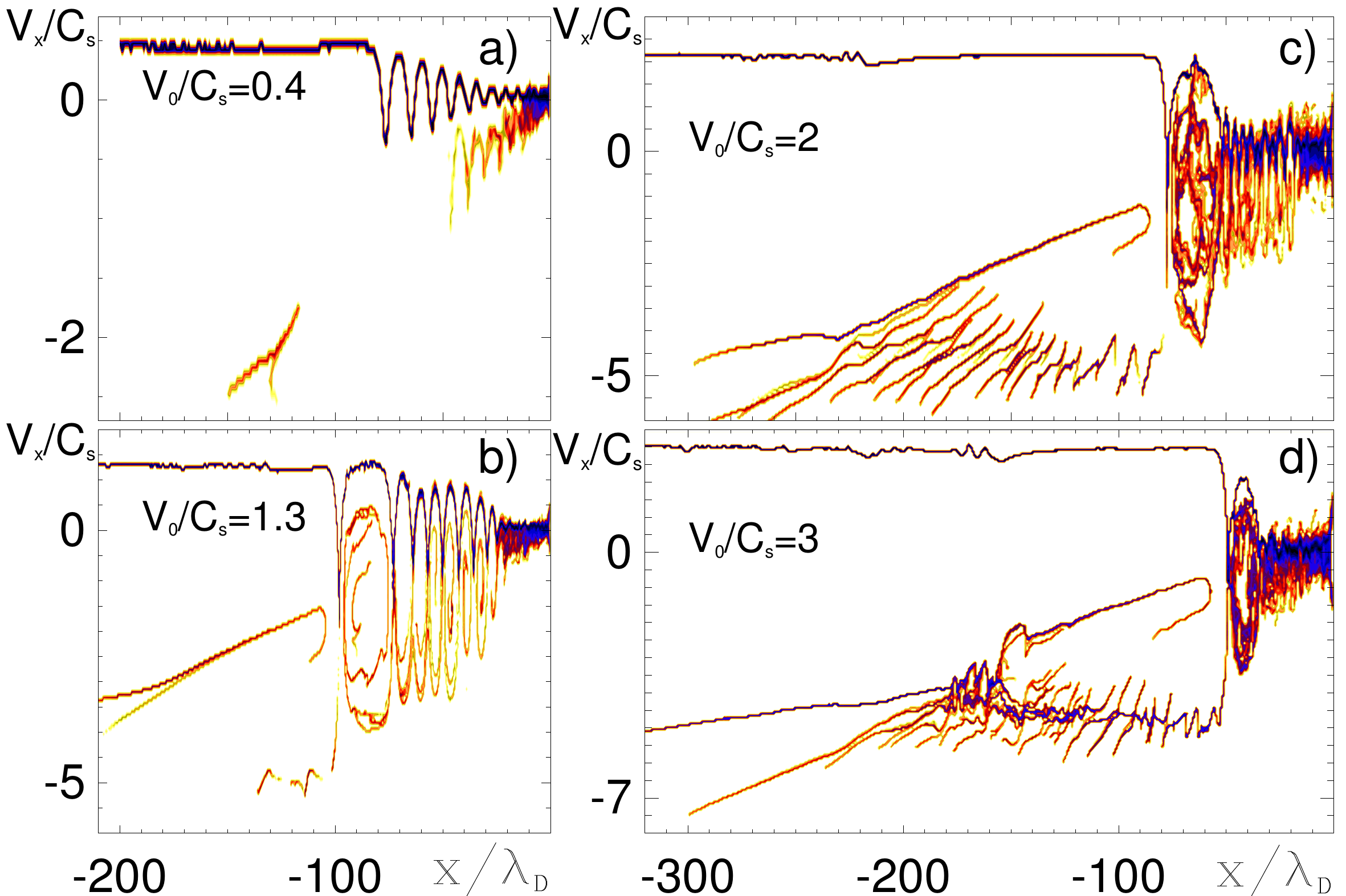} 
\includegraphics[width=0.43\textwidth]{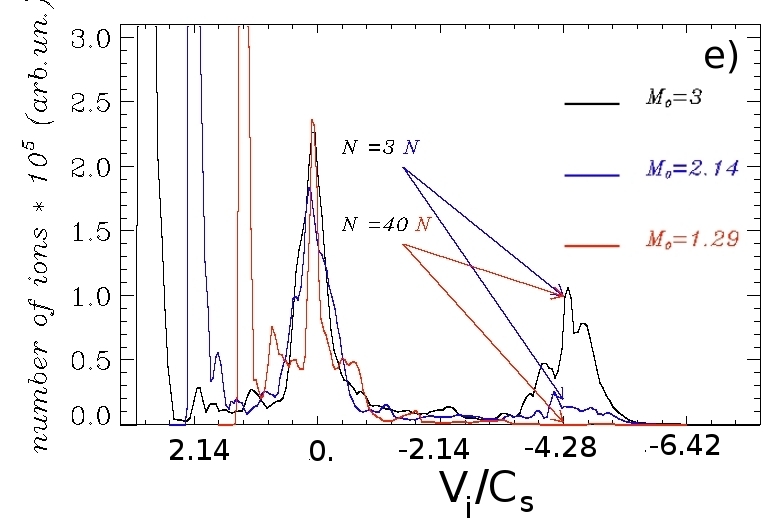}
\caption{\textit{a)--d) -- Distribution of ions on the $(X, V_x)$ phase plane for different values of $M_0=V_0/C_s.$ e) -- 
Distribution function of ions $f_i(v_i).$}}
\label{fig:v0=diff}
\end{figure*}

The number of ions reflected out of the shock increases with the increasing of the flow velocity, see Fig. \ref{fig:v0=diff}, where 
the distributions of ions on the phase plane $(X,V_x)$ and the distribution functions of ions $f(v)$ at $t\omega_{pe}=2800$ are shown. 
The distribution function has three distinct peaks: inflowing ions ($v = V_0$), reflected ions ($\lvert v \rvert \simeq (V_0+2 V_{front}),$ 
with $V_{front}$ the speed of the reflecting hump), and ions at rest downstream of the shock. The energy spectrum of the reflected ions
is very broad.
%, the one on the left corresponds to the incoming flow, the two additional peaks
%are related to the reflected ions ($v\gt 2V_0/C_s$), almost all the ions in the second peak ($v > V_0+2 V_{front})/C_s$) are the ions reflected from the shock.

\begin{figure*}
\centerline{
\includegraphics[width=0.47\textwidth]{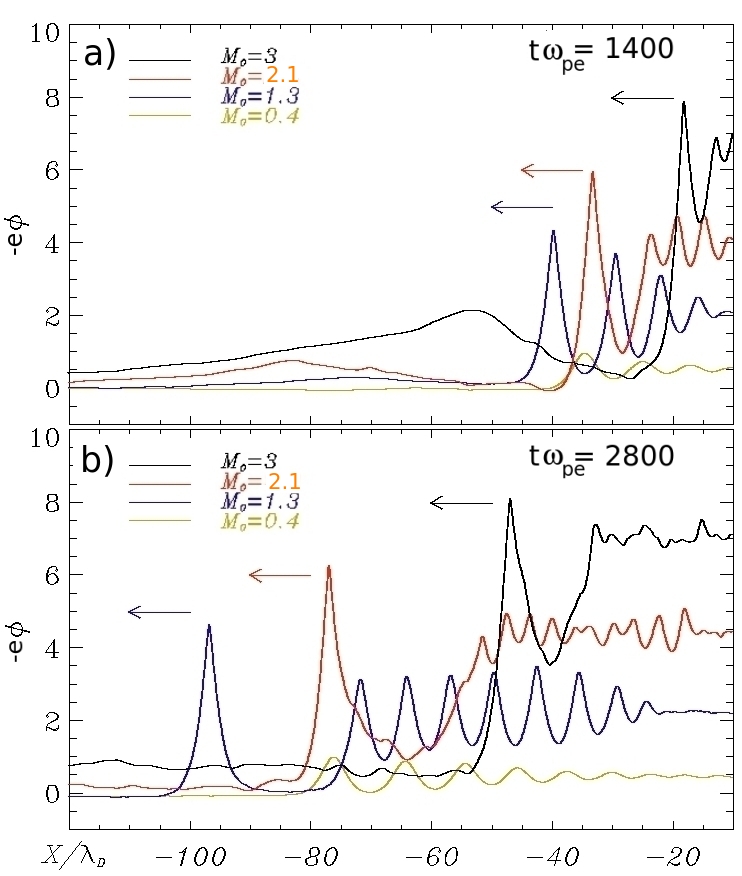}
\includegraphics[width=0.51\textwidth]{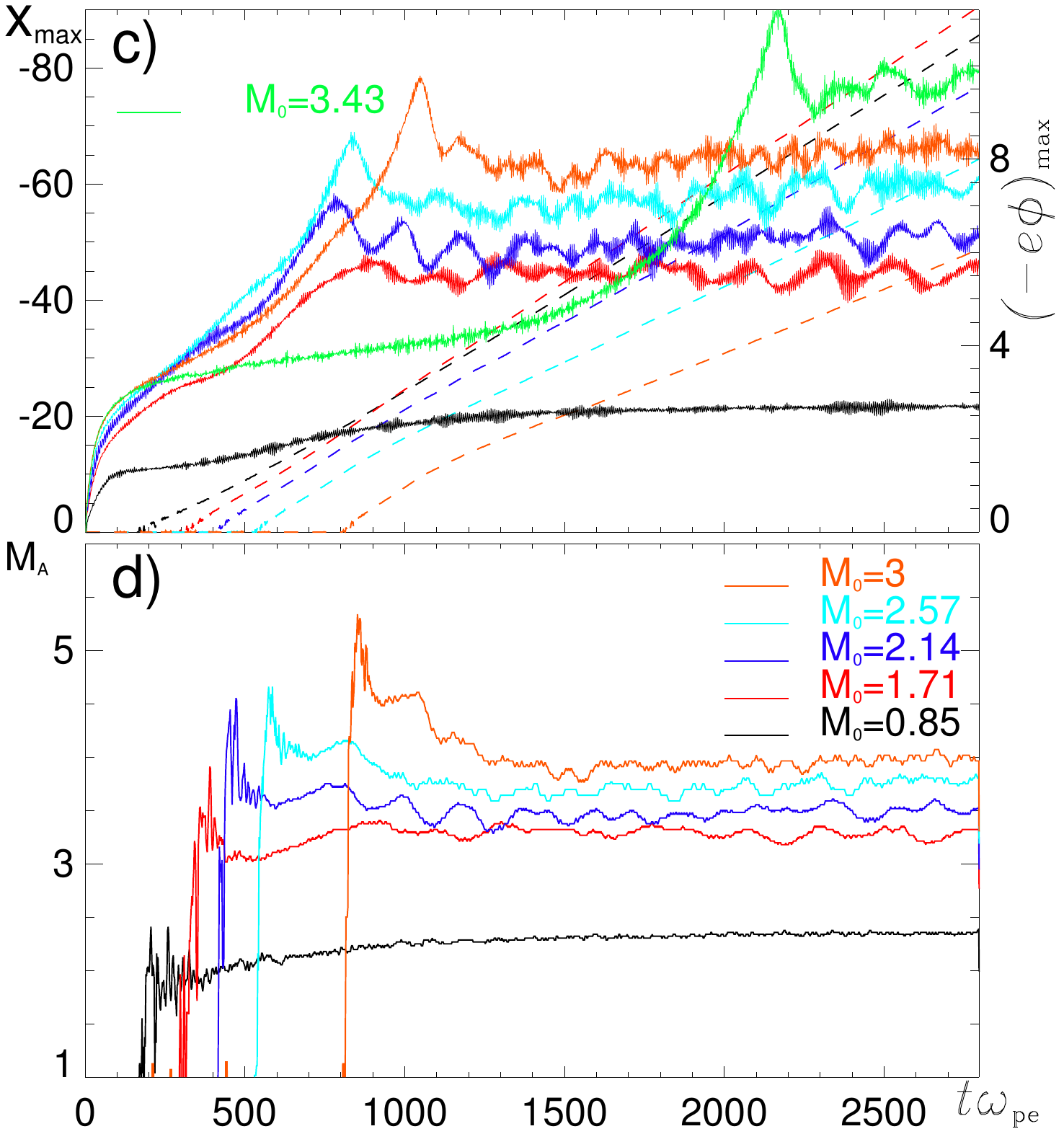}
}
\caption{\textit{a)--b) -- Electrostatic potential for $M_0=0.4,1.3,2.1$ and $3.$ c)--d) -- Evolution of the maximum of the electrostatic potential and the velocity of the shock wave in time. The cusp structure in the potential occurs 
if cold ions are reflected. The outer soliton is liberated from its dispersive tail. The solution is asymmetric with respect 
to the reflection point, its \textit{downstream} is oscillatory. \textit{Upstream} of the soliton, reflected ions create 
a foot with an elevated electrostatic potential.}}
\label{fig:phi_v0=diff}
\end{figure*}

The spatial distributions of the electrostatic potential ($-e\phi$) for different values of upstream flow are presented in 
Fig. \ref{fig:phi_v0=diff} a)--b). It is clearly seen that
%The cusp structure in the potential for  $M_0=1.3, 2.3, 3$ occurs, because cold ions are reflected.
%The outrunner is a soliton, liberated from its dispersive trail. The solution is asymmetric with respect to the reflection point, 
%its \textit{downstream} is oscillatory. \textit{Upstream} of the soliton, reflected ions create a foot with an elevated electrostatic 
%potential. 
when the ions begin to reflect from the soliton tip, the classical single solution bifurcates 
into a more complex structure that comprises the leading soliton, the periodic wave train downstream of it and the foot occupied by 
the reflected ions. This foot is supported by the reflected ions and also accelerates them somewhat further. 
%During the expansion into the 
%upstream medium the reflected ions cool down significantly. 
Because of the reflection of cold ions the leading soliton for $M_0=1.3,2.1$ and $M_0=3$ acquire a clear cusp structure. 

The particle reflection strongly depends of the amplitude of the shock (overshoot of electrostatic potential). In turn it alters the 
shock amplitude and speed, and thus the reflection threshold. Moreover, the reflected ions perturb the electrostatic potential upstream 
which changes the speed of inflowing ions and thus, again, the condition for the ensuing reflection. For $M_0=2.1$ and $M_0=3$ 
the shock looses much of its momentum because of the ion reflection, for $M_0=1.3$ the number of reflected ions is very small. Momentum 
conservation suggests that ion reflection slows down the shock, thus decreasing its amplitude. Evidently, decreasing amplitude acts against 
reflection. However, if the amplitude drops below the reflection threshold for a given speed, 
%(that, in turn, can not drop below speed of sound) 
the shock amplitude 
should recover and the reflection will resume. By virtue of these feedback loops the reflection efficiency can be expected to either stay at a 
self-organized critical level or to exhibit cyclic dynamics. Numerical simulations and observations
validate both scenarios, though, the cyclic shock behavior was primarily observed in magnetized
plasma environments \cite{cyclic,cyclic1}. The temporal evolution of the maximum value ($-e\phi_{max}$) of the electrostatic potential 
extracted from our simulations, Fig. \ref{fig:phi_v0=diff} c) (solid line), confirms a cyclic behavior of the shock.\\ 
%Nevertheless, it is logical to begin investigation of shocks that are formed entirely due to ion
%reflection by seeking such stationary shock transition. 
\indent The approximate velocity of the shock with respect to the upstream flow can be defined from the position of the maximum of the 
electrostatic potential (dashed curves in Fig. \ref{fig:phi_v0=diff} c)). The result is shown in Fig. \ref{fig:phi_v0=diff} d). 
Measured in units of $C_s,$ 
this velocity directly represents the shock Mach number $M_A=V_{shock}/C_s.$ Our simulations prove that the solution, consisting of the leading soliton, 
periodic wave train downstream and the foot occupied by the reflected ions persists with the increasing Mach numbers up to 
a critical value of $M_A=M_2\simeq 4.5$ until almost all incident ions get reflected. In case of cold ions this critical value is 
reached for the upstream velocity $V_0=V_2=3.43 C_s.$ The obtained critical value of $M_2$ is much beyond a critical 
value $M_{\star}$, at which the shock is about to reflect some of the upstream ions. For the cold upstream ions and Bolzmannian electrons 
(this is definitely not the case in our simulations) $M_{\star}\simeq 1.6$ \cite{Sagdeev}, for adiabatically trapped 
electrons $M_{\star}\simeq 3.1$ \cite{Gurevich}. Note, that in the early 
days of collisionless shock research the existence of stationary shock transitions for Mach numbers exceeding critical $M_{\star}$ has 
been often disfavored. 
\begin{figure}
\centerline{
\includegraphics[width=0.5\textwidth]{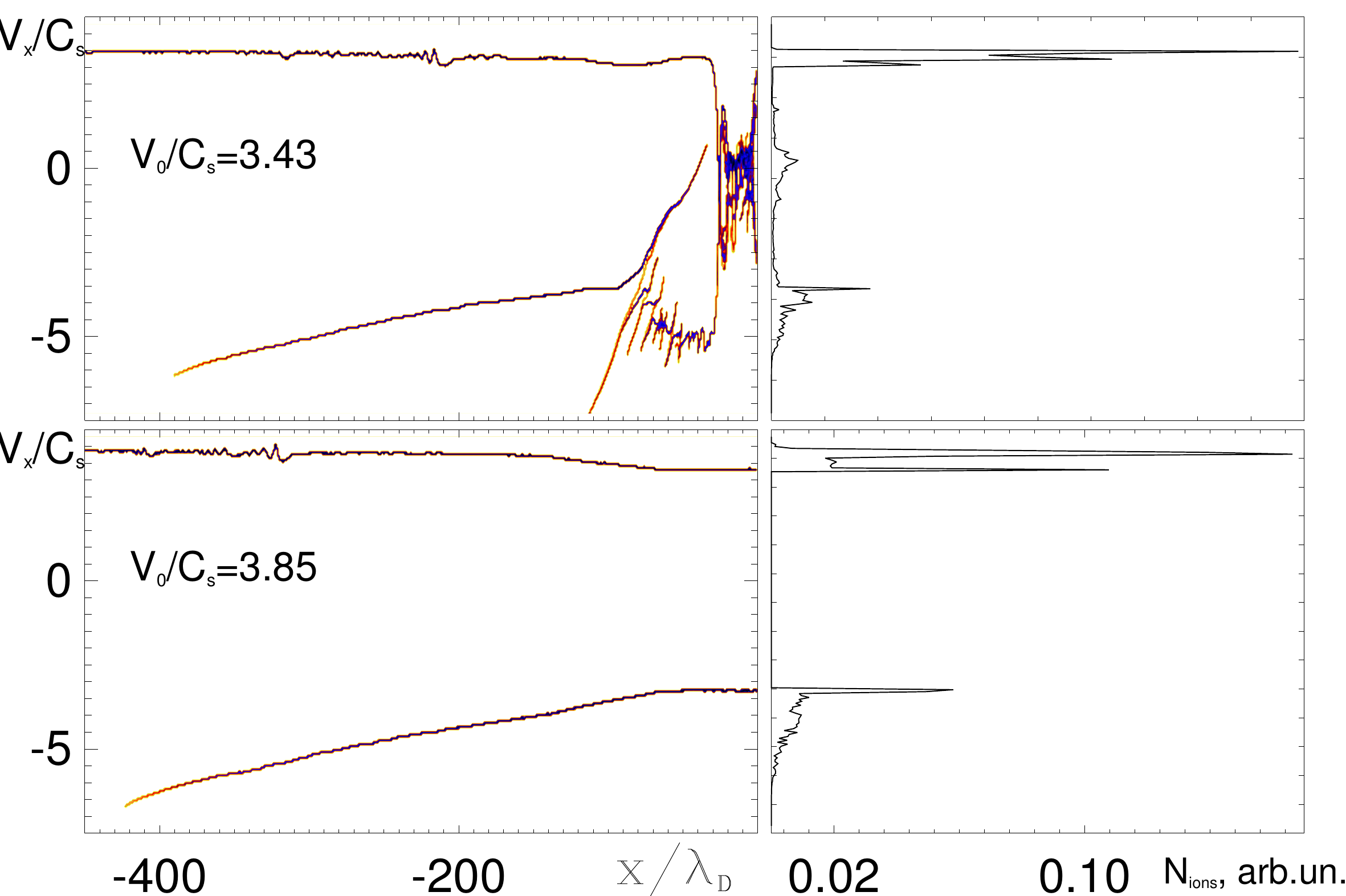}}
\caption{\textit{(color online) Distribution of ions on $(X,V_x)$ phase plane (left frames) and
distribution function of ions $f_i(v_i)$ (right frames) at $t\omega_{pe}=2800$ for $M_0=3.43$ and $M_0=3.85.$ For $M_0=3.85$ the
flow velocity of the incident plasma exceeds the critical value $V_2.$}} 
\label{fig:bigMa}
\end{figure}
 
When the velocity of the incoming flow is bigger than $V_2$ the solution becomes qualitatively very different. 
An almost monoenergetic beam of ions, produced by the reflection of the inflowing ions with 
$M_0=3.85,$ is clearly seen in Fig.\ref{fig:bigMa}. Note also the sharp peak in the distribution function ($v_i<0$) 
and the significant deceleration of the inflowing ions ($v_i>0$)
for $M_0=3.85.$
The next qualitative change in the solution should occur when the inflow velocity reaches the values comparable to the thermal velocity of 
the electrons. At this point the Buneman instability starts to play an important role in the interaction of flows. This regime is 
however outside the current study.

\begin{figure*}
\centerline{
\includegraphics[width=0.8\textwidth]{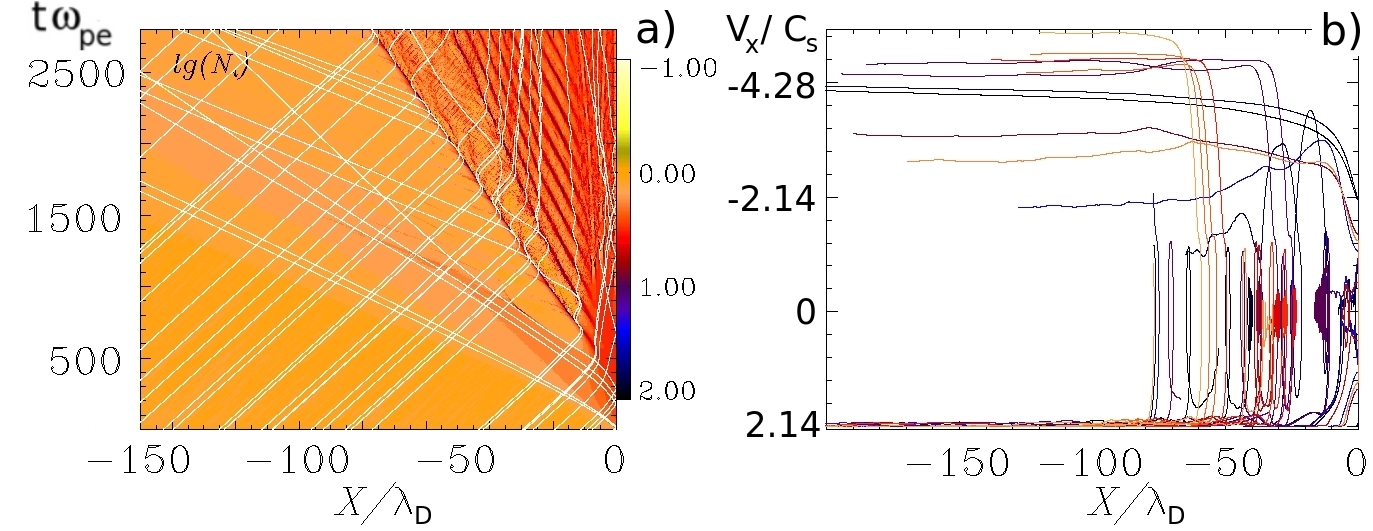}}
\caption{\textit{a) -- Ion density distribution and ion time history for $M_0=2.14$ b) -- Particles trajectories on $(X(t),V_x(t))$ plane.}}
\label{fig:rho}
\end{figure*}

Finally, frame a) of Fig. \ref{fig:rho} shows the distribution of ion density $n_i(x,t)$ in logarithmic scale and the time history $x_i(t)$ of some 
selected ions. The first peak in the ion density corresponds to the ``outrunner''  soliton, liberated from its dispersive trail. 
The velocity of this soliton is almost constant in time (apart from the oscillations, mentioned before) and it reflects a lot of incoming 
ions, which then escape with double the shock speed, see also Fig. \ref{fig:rho} b). The trail, instead, not only propagates with a smaller 
velocity, it slows down significantly as well. Some of the ions get trapped between the soliton and the leading part of the trail.

\begin{figure*}
\includegraphics[width=0.8\textwidth]{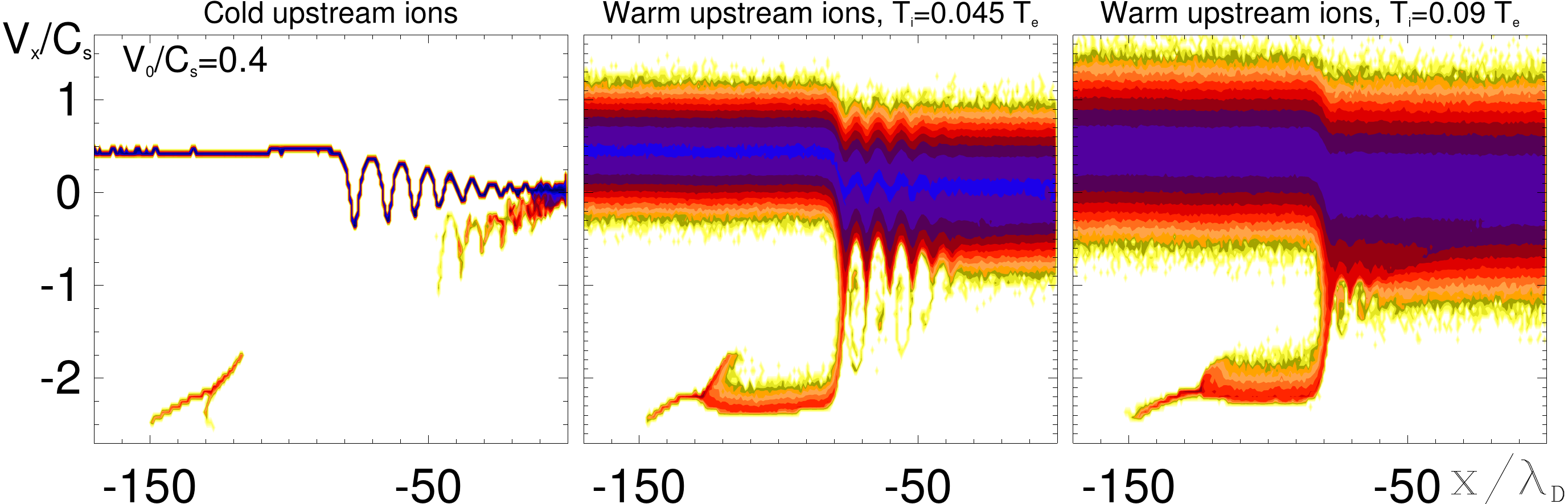}
\caption{\textit{Distribution of the ions on $(X, V_x)$ phase plane for $M_0=V_0/C_s=0.4$ and $T_i=0$ (left frame), 
$T_i=0.045 T_e$ (middle frame), $T_i=0.09 T_e$ (right frame).}}
\label{fig:temp}
\end{figure*}

\subsection{Warm upstream ions}
The resolution of our PIC simulations is high enough to accurately include the finite ion temperature in our analysis. 
Fig.\ref{fig:temp} shows the distributions of ions on the phase plane for $M_0=0.4$ (very slow upstream flow) 
and cold/warm upstream ions.  
If the upstream ions are cold ($T_i=0$) only a negligibly small number of them get reflected,
but if $T_i > 0$ the final distribution function of ions assumes a different shape. In Fig.\ref{fig:temp} 
a \textit{steady reflection} of ions is clearly visible for $T_i>0.$ Their number remains, however, very small and 
the energy spectrum -- quite broad. Moreover, there are ions trapped in the ``multipeaks''.  
Note, that stability of solitons and formation of the electrostatic shock waves in  
ultraintense laser interaction with overdense plasmas  are also found to be strongly dependent
on the velocity distribution of ions\cite{MacchiPRE}. 
It should be mentioned, that if the flow velocity of warm (both electrons and ions) inflowing plasma exceeds 
$V_2$ the solution qualitatively similar to Fig.\ref{fig:bigMa} forms again: the reflection produces a huge amount of almost 
monoenergetic ions which propagate away from the shock.

\section{Conclusion}
Our simulations show that (i) when the Mach number of the acoustic soliton increases to the point of ion  
reflection and the \textit{soliton} transforms into a \textit{soliton train} downstream, this structure persists with 
increasing Mach number until almost all incident ions get reflected; (ii) the ion reflection alters the shock amplitude 
and speed, thus impacting the reflection threshold itself; (iii) the pedestal of the electrostatic 
potential, supported by the reflected ions, changes the speed of inflowing ions and thus again, the condition for the ensuing reflection.
The formation of the pedestal in the potential between the reflecting hump and the front edge of the group of reflected ions 
makes the analytic theory very challenging. In fact, this pedestal is sensitive
to the model used for the transition between the reflected ions and upstream (e.g. cold vs hot beam of reflected ions).
In the pedestal the inflowing plasma is slowed  down considerably, if its front edge moves at the speed of the main shock, 
but reflected particles pass through it and accelerate, decreasing their density. 
%What kind of electric potential is associated with reflected particles when they run way ahead of 
%the reflecting hump? 
All these uncertainties are critical and hard to resolve without simulations. 
Note, that an analytic solution for the ion-acoustic collisionless shock with self-consistently reflected ions and 
their subsequent cooling was recently presented \cite{IAshock13,IAshock14}. Its relation to the well known non-reflecting 
ion-acoustic soliton solution, strictly limited by the critical Mach number $M_{\star}= 1.6$ (Boltzmann electrons)\cite{Sagdeev} 
and  $M_{\star}= 3.1$ (trapped electrons) \cite{Gurevich}, was clarified and quantified.

This work was in part supported by the DFG within the SFB 652 and by RSCF under the Grant 14-11-00485. 
PIC simulations were performed using the computing resources granted by the John von Neumann-Institut for Computing 
(Research Center J\"{u}lich) under the project HRO03.

\end{document}